\documentclass[prb,twocolumn,showpacs,preprintnumbers,amsmath,amssymb]{revtex4-1}

\usepackage{graphicx}
\usepackage{bm}
\usepackage{epstopdf}
\usepackage{color}
\usepackage{cancel}

\begin{document}

\title{Linking Magnon-Cavity Strong Coupling to Magnon-Polaritons through Effective Permeability}
\author{Paul Hyde}
\email{umhydep@myumanitoba.ca}
\author{Lihui Bai}
\email{bai@physics.umanitoba.ca}
\author{Michael Harder}
\author{Christopher Dyck}
\author{Can-Ming Hu}

\affiliation{Department of Physics and Astronomy, University of Manitoba, Winnipeg, Canada R3T 2N2}

\date{\today}

\begin{abstract}

Strong coupling in cavity-magnon systems has shown great potential for use in spintronics and information processing technologies due to the low damping rates and long coherence times. Although such systems are conceptually similar to those coupled by magnon-polaritons (MPs), the link between magnon-cavity coupling and MPs has not been explicitly defined. In this work we establish such a connection by studying the frequency-wavevector dispersion of a strongly coupled magnon-cavity system, using a height-adjustable microwave cavity, and by modelling the observed behaviour through the system's effective permeability. A polariton gap between the upper and lower coupled modes of the magnon-cavity system is defined, and is seen to be dependent on the system's effective filling factor. This gap is equal to the MP polariton gap in the limit where filling factor = 1, corresponding to the removal of the microwave cavity. Thus, our work clarifies the connection between magnon-cavity and MP coupling, improving our understanding of magnon-photon interactions in coupled systems.

\end{abstract}

\maketitle

\section{Introduction}

Strong coupling at room temperature between magnon and photon systems inside of a cavity resonator has attracted a lot of attention recently\cite{A,B,C,D,E,F,G}, due to its significant potential for use in new information processing technologies\cite{H,I,II}. These systems typically place a magnetic material inside of a microwave cavity and control its resonant frequency through the application of an external magnetic field. By tuning the magnon resonance frequency of the material to approach the cavity resonance frequency, coupling between these two systems can be detected in frequency-magnetic field ($\omega$-H) dispersions both optically and electrically. To describe the effects of coupling in these systems many models have been developed, including a dynamic circuit model\cite{F}, transfer matrices model\cite{G,N}, and coupled harmonic oscillator model\cite{A,C,F,KK}. Major features of strong coupling in both measured and modelled $\omega$-H dispersions include mode anti-crossing, damping evolution, and phase induced line shape changes.

Due to the similarity to how magnon-polaritons (MPs) couple electromagnetic (EM) fields to magnon excitations in magnetic materials, the coupling in magnon-cavity systems is commonly described as being controlled by quasi-particles known as cavity magnon-polaritons (CMPs) \cite{N,P,R}. MP coupling is typically measured at optical frequencies in frequency-wavevector ($\omega$-k) dispersions; these dispersions display a characteristic frequency gap between resonant modes, known as a polariton gap \cite{K,L,M}. In coupled magnon-cavity systems measurements are generally taken as frequency-magnetic field ($\omega$-H) dispersions, which cannot display a polariton gap. Since a CMP coupled system is expected to share many key dispersion features with MP coupling, the inability to measure this characteristic feature of MP coupling in magnon-cavity systems leaves the physical connection between the coupling mechanisms of each system unclear.

In this work we use a height-tunable cavity to measure the $\omega$-k dispersion of a coupled magnon-cavity system. These measurements are used to develop a model which calculates the effective permeability of the coupled system by approximating the filling factor of the magnetic material within the cavity. The measurements and model both reveal the existence of a polariton gap produced by coupling in the magnon-cavity system, with the magnitude of the gap found to be strongly dependent on filling factor. The presence of this gap in magnon-cavity systems establishes a connection between MP and magnon-cavity-based coupling features, and shows that the two systems share a common means of coupling. By using a single model to link MP and magnon-cavity-based coupling, our work expands the understanding of magnon-photon interactions in magnon-cavity systems, and should be useful for future information processing technology designs based on magnon-photon coupling. After developing our model in Section II, a brief discussion of the components of our magnon-cavity system is given in Section III. Using a height adjustable cavity, we fit our model to measured $\omega$-k$_{z}$ dispersions in Section IV(A), and measure material filling factors roughly in agreement with the approximated values. In Section IV(B), our model is fit to measured $\omega$-H dispersions and it is shown that the results produced agree with those of previously used models\cite{A,F,G,N,C,KK}. In Section V we compare the relations the $\omega$-k$_{z}$ polariton frequency gap and the $\omega$-H Rabi coupling gap have on the material filling factor of our system and show that, through our model, magnon-cavity coupling effects can be characterized by the magnitude of these gaps. Finally, in Section VI we summarize the results presented in this paper.

\section{Theory}

\begin{figure*}[t]
\centering
\includegraphics[width=130mm]{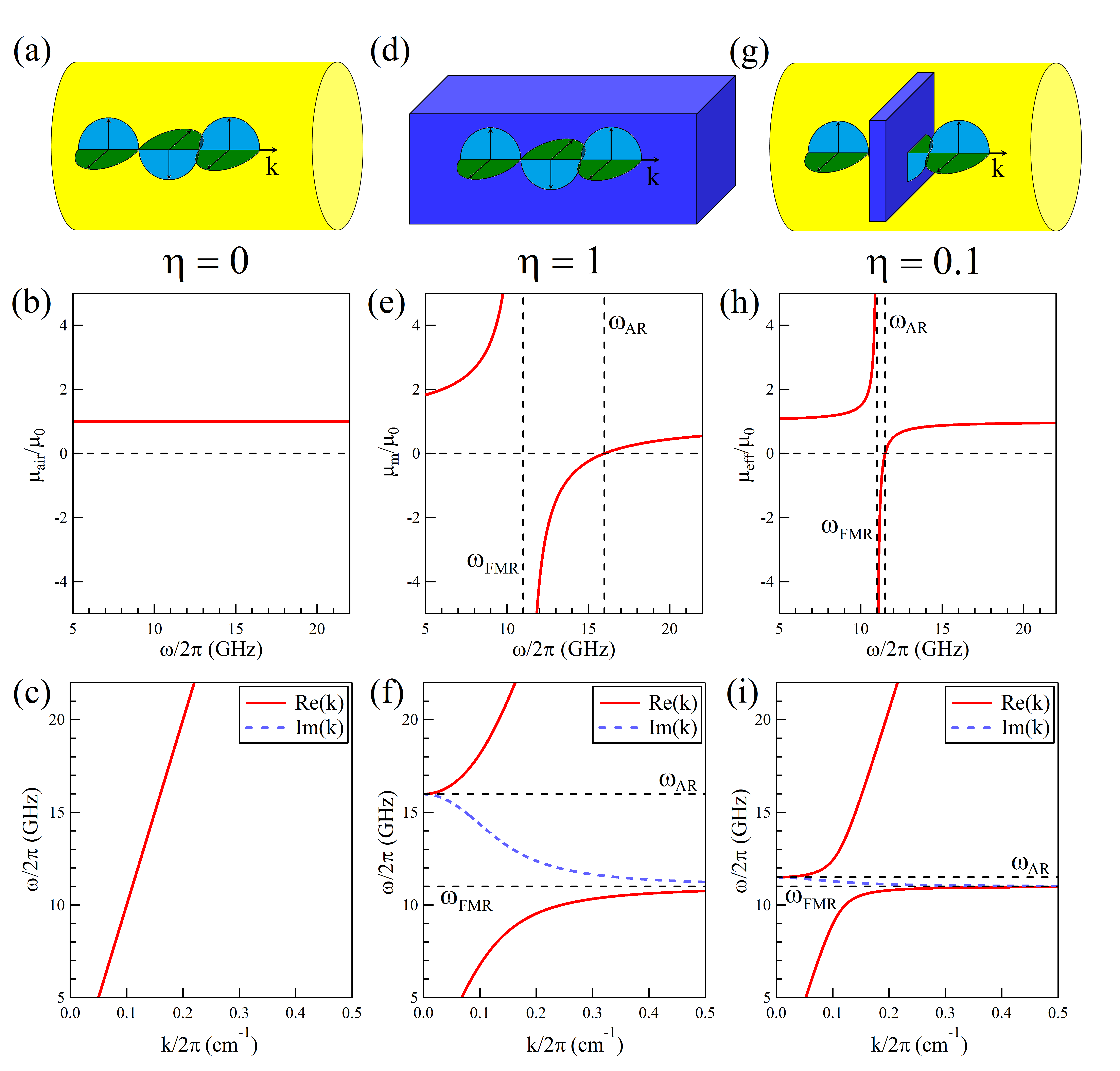}
\caption{(Colour online) \textbf{(a)} In an empty cavity [filling factor ($\eta$) = 0 in Eq. 1] the permeability, $\mu_{air}$ of the air inside \textbf{(b)} is roughly independent of $\omega$. \textbf{(c)} The EM wave vectors, k, able to propagate through the cavity thus increase linearly with $\omega$. \textbf{(d)} For the case where the EM wave is moving through a magnetic material ($\eta=1$ in Eq. 1) the material permeability, $\mu_{m}$ exhibits a strong dependence on $\omega$. \textbf{(e)} At $\omega_{FMR}$ the value of $\mu_{m}$ diverges and at a higher frequency, $\omega_{AR}$, crosses zero from negative to positive values. \textbf{(f)} The change in $\mu_{m}$ results in the wave vector, k, approaching $\infty$ at $\omega_{FMR}$ and reaching zero at $\omega_{AR}$. Between these two values $\mu_{m}$ becomes negative, resulting in a purely imaginary k value. \textbf{(g)} For an EM wave travelling in a cavity partially filled with both air and magnetic material, we can use Eq. 1 to approximate the filling factor of the material. \textbf{(h)} For $\eta=0.1$ we see the effective permeability of the cavity/material system, $\mu_{eff}$ behaves similar to $\mu_{m}$ in (e), however the space between $\omega_{FMR}$ and $\omega_{AR}$ is decreased by a factor of $\eta$. \textbf{(i)} The k vector again approaches $\infty$ at $\omega_{FMR}$ and 0 at $\omega_{AR}$, but similar to (h) the distance between these values is reduced by a factor of $\eta$. The imaginary k values between $\omega_{FMR}$ and $\omega_{AR}$ in (f) and (i) prevent microwaves from passing through the material, meaning no system resonance modes can be observed between these frequencies.}
\label{Figure 1}
\end{figure*}

The behaviour of electric and magnetic fields in materials is governed by Maxwell's equations, which state that the wave vector, k, of an electromagnetic wave, $h_{em}$, is described by the relation: $[k^{2}-\omega^{2}\epsilon(\omega)\mu(\omega)]h_{em}=0$. Here we see that the propagation of the wave is directly dependent on its frequency, $\omega$, as well as the permittivity, $\epsilon$, and permeability, $\mu$, of the medium it is travelling through. If the wave is travelling through air [as sketched in Fig 1(a)] both $\epsilon$ and $\mu$ will be independent of $\omega$, resulting in the linear $\omega$-k dispersion, $k^{2}=\omega^{2}\epsilon\mu_{0}$, shown in Fig. 1(c).

If the EM wave is travelling through a magetic material, as in Fig. 1(d), $\epsilon$ will remain constant but $\mu$ will become frequency dependent. Assuming a configuration where k is parallel to the material's polarization, the FM permeability can be expressed as $\mu_{m}=1+\chi_{L}+i\chi_{T}$, where $\chi_{L}=\omega_{FMR}\omega_{m}/(\omega_{FMR}^{2}-\omega^{2})$ and $\chi_{T}=-i\omega\omega_{m}/(\omega_{FMR}^{2}-\omega^{2})$ are the longitudinal and transverse elements of the Polder tensor resulting from the LLG equation (for small damping)\cite{MM}. Here, $\omega_{m}=\gamma M_{0}$ is the saturation frequency, determined by the gyromagnetic ratio, $\gamma$, and the saturation magnetization, M$_{0}$, of the material. Looking at the expression for $\mu_{m}$ as frequency, $\omega$, is changed in Fig 1(e), we see several important features. At $\omega=\omega_{FMR}$, $\mu_{m}$ approaches infinity, indicating that the material is completely absorbing the incident wave. This behaviour is well studied as ferromagnetic resonance (FMR), and the frequency it occurs at, $\omega_{FMR}=\gamma H$, is determined by the strength of an external magnetic field, \textbf{H}, acting on the material. At a higher frequency, $\omega_{AR}=\omega_{FMR}+\omega_{m}$, the permeability of the material will equal zero, resulting in behaviour known as ferromagnetic anti-resonance (FMAR).

In Fig 1(e) we see that between $\omega_{FMR}$ and $\omega_{AR}$, $\mu_{m}$ becomes negative. Plotting the propagation of the incident wave vector through the material, $k^{2}=\omega^{2}\epsilon\mu_{0}\mu_{m}$, in Fig 1(f), we see that as $\omega$ approaches $\omega_{FMR}$ the real value of the wave vector, Re(k), increases to infinity. As $\omega$ approaches $\omega_{AR}$, Re(k) is reduced to zero. Between these frequencies, where $\mu_{m}$ is negative, k will have a purely imaginary value. This value, Im(k), induces a large amplitude decay within the material as the wave attempts to pass through, effectively blocking the transmission of electromagnetic waves at all frequencies and leaving a visible frequency gap in the $\omega$-k dispersion. This frequency gap is a result of magnon-polaritons being created as the electromagnetic wave interacts with the material, and is called a polariton gap.

If, instead of having our EM wave travel through a homogeneous medium, we confine it to a microwave cavity partially filled with magnetic material [Fig. 1(g)], the exact solutions to Maxwell's equations will depend on the field distribution, $h_{em}$, within the system. This field distribution can be used to determine the filling factor of the magnetic material, $\eta$, which is calculated as the total magnetic energy stored in the material as a fraction of the total magnetic energy of the system, $\eta=\int_{V_{m}}|h_{em}|dV/\int_{V_{tot}}|h_{em}|dV$. In general, mode dependent field distributions within both the cavity and material systems will make calculating $\eta$ difficult. However, if we assume a homogeneous field distribution throughout the system, this filling factor can be approximated as the volume ratio of the system components, $\eta\simeq V_{m}/V_{air}$, allowing us to define an effective permeability, $\mu_{eff}$, for the system as, $\mu_{eff}=\mu_{air}(V_{air}-V_{m})/V_{air}+\mu_{m}(V_{m}/V_{air})$. Here $\mu_{air}$ and $V_{air}$ are the permeability and volume of the air filled section of our system, and $\mu_{m}$ and $V_{m}$ are the permeability and volume of the material. Taking $\mu_{air}=\mu_{0}$, we can write the effective permeability of the system as $\mu_{eff}=1-\eta+\eta\mu_{m}$. Plotting $\mu_{eff}$ as a function of k in Fig. 1(h) for $\eta=0.1$ we see that it behaves similarly to $\mu_{m}$ in Fig. 1(e), diverging at $\omega_{FMR}$ and crossing zero at $\omega_{AR}$. However, the distance between $\omega_{FMR}$ and $\omega_{AR}$ has been reduced here by a factor of 0.1. Applying this effective permeability, the microwave dispersion in the system will be expressed as:

\begin{equation}
k^{2}=\omega^{2}\epsilon\mu_{0}(1-\eta+\eta\mu_{m})
\end{equation}

For the case where $\eta=0$, representing an empty cavity, we see that Eq. 1 will reduce to the dispersion equation for an EM wave in free-space. For the case where $\eta=1$, representing a cavity filled with material, we retrieve the relation describing MP coupling. Plotting Eq. 1 for $\eta=0.1$ in Fig. 1(i) we see the wave vector, k, behaves similar to Fig. 1(f), approaching infinity at $\omega_{FMR}$ and becoming zero at $\omega_{AR}$. Between these two frequencies the value of $\mu_{eff}$ is negative and k is entirely imaginary. In this range there will be no detectable resonant modes within the system, meaning that $\omega$-k dispersion measurements will see a gap between modes at frequencies below $\omega_{FMR}$ and those above $\omega_{AR}$. The presence of this frequency gap between resonant modes is a well-known consequence of MP coupling. Using the effective permeability of a coupled material/cavity system to model the effects of magnon-cavity coupling, we reveal that although the magnitude of this polariton gap will be reduced, it remains detectable within these systems.

\section{Experiment}

To produce magnon-cavity coupling we place a Yttrium Iron Garnet (YIG) sphere\cite{S} inside of a cylindrical microwave cavity composed of oxygen-free copper. The sphere had a diameter of 1 mm and was positioned on the bottom of the cavity as shown in the inset of Fig. 2. The YIG had a saturation frequency $\omega_{m}$ = $2\pi\times$4.984 GHz, and a Gilbert damping coefficient $\alpha$ = 1.5 $\times$10$^{-4}$. Since $\alpha\ll 1$ we can assume the effects of damping will be small, allowing us to ignore damping terms in our calculations. Interactions between the YIG and the microwave fields within the cavity can be controlled by applying an external magnetic field, \textbf{H}, to the system, creating a magnon-cavity coupled system. FMR in the YIG (when not coupled to the cavity) occurs at a frequency of $\omega_{FMR}=2\pi\times\gamma(H+H_{a})$, where the gyromagnetic ratio of YIG is $\gamma=176\mu_{0}$ GHz/T and the anisotropy field of the sphere is $\mu_{0}H_{a}$ = -2.4 mT.

To measure magnon-cavity coupling over a range of wavevector, k, values, the microwave cavity used in our experiment was designed to have a height-tunable cylindrical structure, as shown in the inset of Fig. 2(a). The resonant frequency of the modes produced will be determined by the radius, R = 12.5 mm, and height, h = 25 - 45 mm, of the cavity, with the resonant frequency of the TM$_{011}$ mode being\cite{T} $\omega_{cavity}/2\pi=(1/\sqrt{\epsilon\mu_{0}})\sqrt{(X_{01}/R)^{2}+(\pi/h)^{2}}=(1/\sqrt{\epsilon\mu_{0}})\sqrt{(k_{\perp})^{2}+(k_{z})^{2}}$. Here $X_{01}$ is the first root of the zeroth Bessel function, while k$_{\perp}$ and k$_{z}$ are, respectively, the wave vector components perpendicular and parallel to the axis of the cavity (z). From this equation we see that the k$_{z}$ vector component is directly dependent on the height of the cavity, with $k_{z}=\pi/h$. In our measurements we focus on the TM$_{011}$ cavity mode due to its circular EM field distribution. As we adjust the cavity height in Fig. 2, we see that the measured resonance modes agree with those calculated for the TM$_{011}$ mode, confirming that the excited mode is the TM$_{011}$ mode. In Fig. 2 we can also note that as k$_{z}\rightarrow 0$ ($h\rightarrow\infty$) the resonance frequency of the TM$_{011}$ mode approaches a minimum value, $\omega_{cutoff}=k_{\perp}/\sqrt{\epsilon\mu_{0}}=2\pi\times9.186$ GHz. This minimum value is generated by the radial component of the TM$_{011}$ mode, which is independent of cavity height; we see later it plays an important role in modifying the $\omega$-k$_{z}$ dispersion of our system during magnon-cavity coupling. In our experimental system, changing the height of the cavity does not significantly change the strength of the microwave field felt by the YIG sample on the bottom of the cavity.

\begin{figure}[t]
\centering
\includegraphics[width= 7.5 cm]{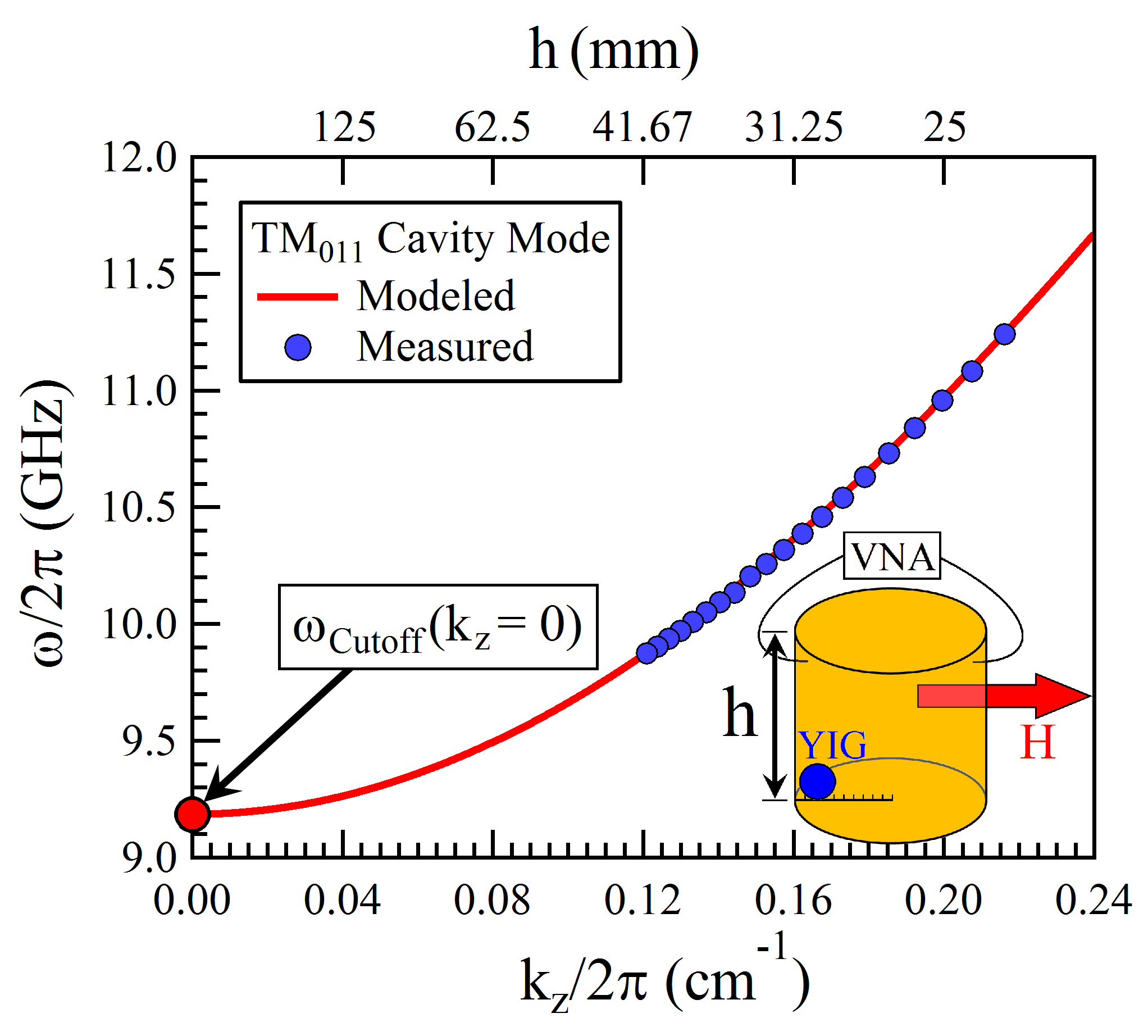}
\caption{(Colour online) A plot of the $\omega$-k$_{z}$ dispersion of the TM$_{011}$ resonance mode within our cavity with no YIG inserted. The solid curve is calculated from Maxwell's equations using the proper boundary conditions, while the markers indicate the measured resonance frequencies of this mode at various cavity height (k$_{z}$) values. Due to the geometry of the cavity, the TM$_{011}$ mode does not equal zero for $k_{z}=0$, but reaches a minimum value indicated by $\omega_{cutoff}$. The inset depicts a diagram of the cavity/YIG set-up used in our measurements.}
\label{Figure 2}
\end{figure}

Since a height-adjustable cavity will only be able to change the value of k$_{z}$ and leave a constant k$_{\perp}$ component at all heights, we are unable to measure a complete $\omega$-k dispersion (from k=0 to k=$\infty$). However, we are able to measure over a sufficiently wide range of k$_{z}$ values to be able to determine whether a frequency gap is produced by magnon-cavity coupling. A key feature of our experiment is the ability to individually change the resonant frequency of both the cavity and YIG systems through changing either cavity height or applied field strength; this allows us to produce both $\omega$-k$_{z}$ and $\omega$-H dispersion plots of our system during coupling.

We examine the effects of magnon-cavity coupling in our cavity/YIG system by measuring microwave transmission, $|$S$_{21}|^{2}$, as a function of cavity height and microwave frequency, producing the $\omega$-k$_{z}$ plot shown in Fig. 3(a). Here an external magnetic field with a constant magnitude of $\mu_{0}$H = 0.4 mT was applied to the system, as shown in the inset of Fig. 2. This field will cause the YIG within the cavity to undergo FMR near $\omega_{FMR}$ = $2\pi\times$10.35 GHz, as indicated by the horizontal dashed line in Fig. 3(a). As the height of the cavity is decreased (k$_{z}$ is increased) the resonant frequency of the TM$_{011}$ cavity mode, $\omega_{Cavity}$ [diagonal dashed line in Fig. 3(a)], gradually decreases towards $\omega_{FMR}$. Near the crossing point of these resonance values the $|$S$_{21}|^{2}$ transmission measurements show the modes anti-crossing with each other, indicating the cavity and YIG systems are coupled with each other. At a frequency somewhat higher than $\omega_{FMR}$ another mode anti-crossing occurs due to the cavity mode interacting with spin wave modes, the effects of which we do not consider here.

\section{Results}

\subsection{$\omega$-k$_{z}$ Dispersion}

The magnon-cavity coupling behaviour seen in the $\omega$-k$_{z}$ dispersion in Fig. 3(a) can be accurately described using the effective permeability model developed in Eq. 1, (solid lines). For this fitting we can determine the cavity and YIG properties from uncoupled cases before their resonance frequencies approach each other, meaning that the only parameter needed when fitting Eq. 1 to Fig. 3(a) is the filling factor $\eta$. In this figure a good fitting is obtained for $\eta=2.3\times 10^{-5}$; this is similar to the ratio between the YIG and cavity volumes (2.7$\times 10^{-5}$ at k$_{z}$ values near coupling) in the system, confirming our previous approximations.

\begin{figure}[t]
\centering
\includegraphics[width= 7.5 cm]{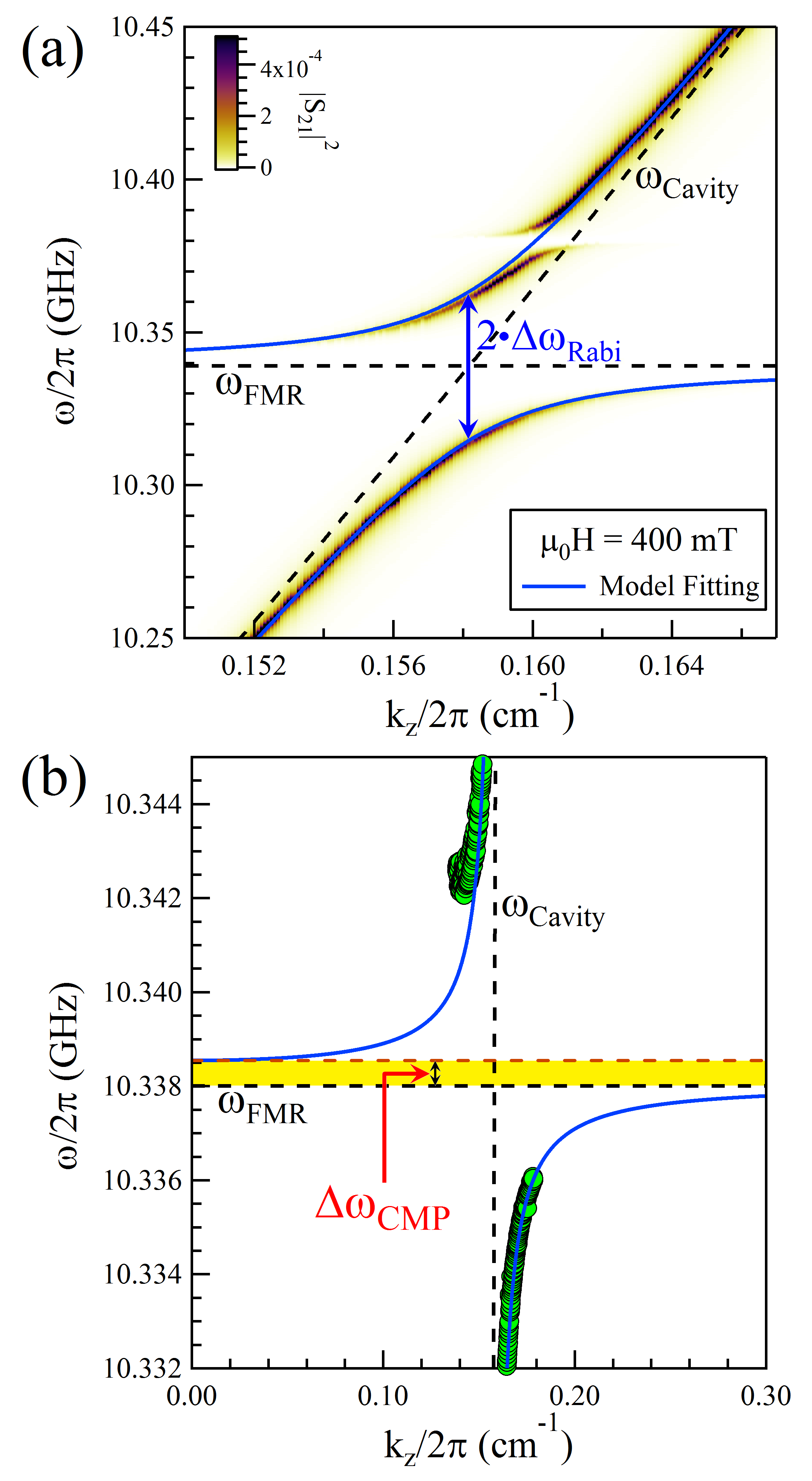}
\caption{(Colour online) \textbf{(a)} An $\omega$-k$_{z}$ plot of the microwave transmission, $|S_{21}|^{2}$, through our cavity/YIG system at an external field strength of 400 mT. The solid blue curve is a fitting of this data to the coupling model in Eq. 1 using $\eta=2.3\times 10^{-5}$, and the black dashed curves indicate the positions of the YIG and cavity resonance modes for the case of no coupling. In this plot the Rabi coupling gap, $\Delta\omega_{Rabi}$, can be measured as half the frequency gap between the upper and lower coupled resonance modes at the k$_{z}$ value where their uncoupled modes would cross. \textbf{(b)} Zooming in this $\omega$-k$_{z}$ plot near $\omega_{FMR}$, we see that in both the measured (markers) and modelled (solid curve) cases the upper and lower coupled resonance modes do not approach the same $\omega$ value as k$_{z}\rightarrow 0$ and k$_{z}\rightarrow\infty$. This leaves a frequency gap, $\Delta\omega_{CMP}$ (highlighted in yellow), where no resonance modes can occur.}
\label{Figure 3}
\end{figure}

From our previous discussion of coupling in magnon-cavity systems, we know that the magnitude of the polariton frequency gap $\Delta\omega_{CMP}$ will be equal to the difference between the YIG FMR and FMAR frequencies. In our cylindrical cavity system these frequencies are reached as k$_{z}\rightarrow\infty$ ($\omega_{FMR}$) and k$_{z}\rightarrow 0$ ($\omega_{AR}$), resulting in the following $\omega$-k$_{z}$ dispersion limits:

\begin{subequations}
\begin{align}
&\omega(k_{z}\rightarrow\infty)& &=& &\omega_{FMR}& \\
&\omega(k_{z}\rightarrow0)& &\simeq& &\omega_{FMR}& + \eta\omega_{m}\frac{1}{1-(\frac{\omega_{Cutoff}}{\omega_{FMR}})^{2}}
\end{align}
\label{eq2}
\end{subequations}

A notable influence on the coupling behaviour of the system is contained within the [1-($\omega_{Cutoff}$/$\omega_{FMR}$)$^{2}$]$^{-1}$ term for the case when k$_{z}\rightarrow$0. This term arises due to the k$_{\perp}$ component of the cylindrical cavity remaining constant as k$_{z}$ goes to zero. The influence of this term can be quite significant when $\omega_{FMR}$ is near $\omega_{Cutoff}$; for the measurements shown in Fig. 3 its value increases the magnitude of $\Delta\omega_{CMP}$ by a factor of 5. For the case where $\omega_{Cutoff}\ll\omega_{FMR}$ ($\omega_{Cutoff}\rightarrow 0$ as cavity radius increases to infinity) this term will reduce to unity and there will no difference between $\omega$-k and $\omega$-k$_{z}$ dispersion plots. As in our magnon-cavity system we have no means of changing k without also affecting the value of $\eta$ we are unable to directly compare our measured results to our Eq. 1 model over a wide range of k values. For the values we are able to measure, however, k (and $\eta$) are shifted by only a relatively small amount, allowing us to approximate $\eta$ as constant within this range. Inserting a constant $\eta$ value into Eq. 1 allows us to fit our model to the measured data in Fig. 3(a) within the measured k range; extending our model past this range while keeping $\eta$ constant reveals the $\Delta\omega_{CMP}$ frequency gap indicative of MP coupling.

Fig. 3(b) shows a magnified plot of Fig. 3(a) at frequencies near $\omega_{FMR}$, at this scale the polariton frequency gap, $\Delta\omega_{CMP}$, created by magnon-cavity coupling is now clearly visible in our modelled results. Due to the small magnitude of $\eta$ for this system ($\eta=2.3\times 10^{-5}$), the size of $\Delta\omega_{CMP}$ is likewise reduced from the MP polariton gap for $\eta=1$ in a magnetic material [Fig. 1(f)]; for the data plotted in Fig. 3 we find $\Delta\omega_{CMP}$ = $2\pi\times$ 0.54 MHz. Our fitted data and the presence of this $\Delta\omega_{CMP}$ polariton gap indicate that Eq. 1 is capable of describing both MP and magnon-cavity coupling, and that the filling factor, $\eta$, of the system is the main factor which controls the transition between these two related modes of coupling.

\subsection{Comparison to $\omega$-H Dispersion}

\begin{figure}[b]
\centering
\includegraphics[width= 7.5 cm]{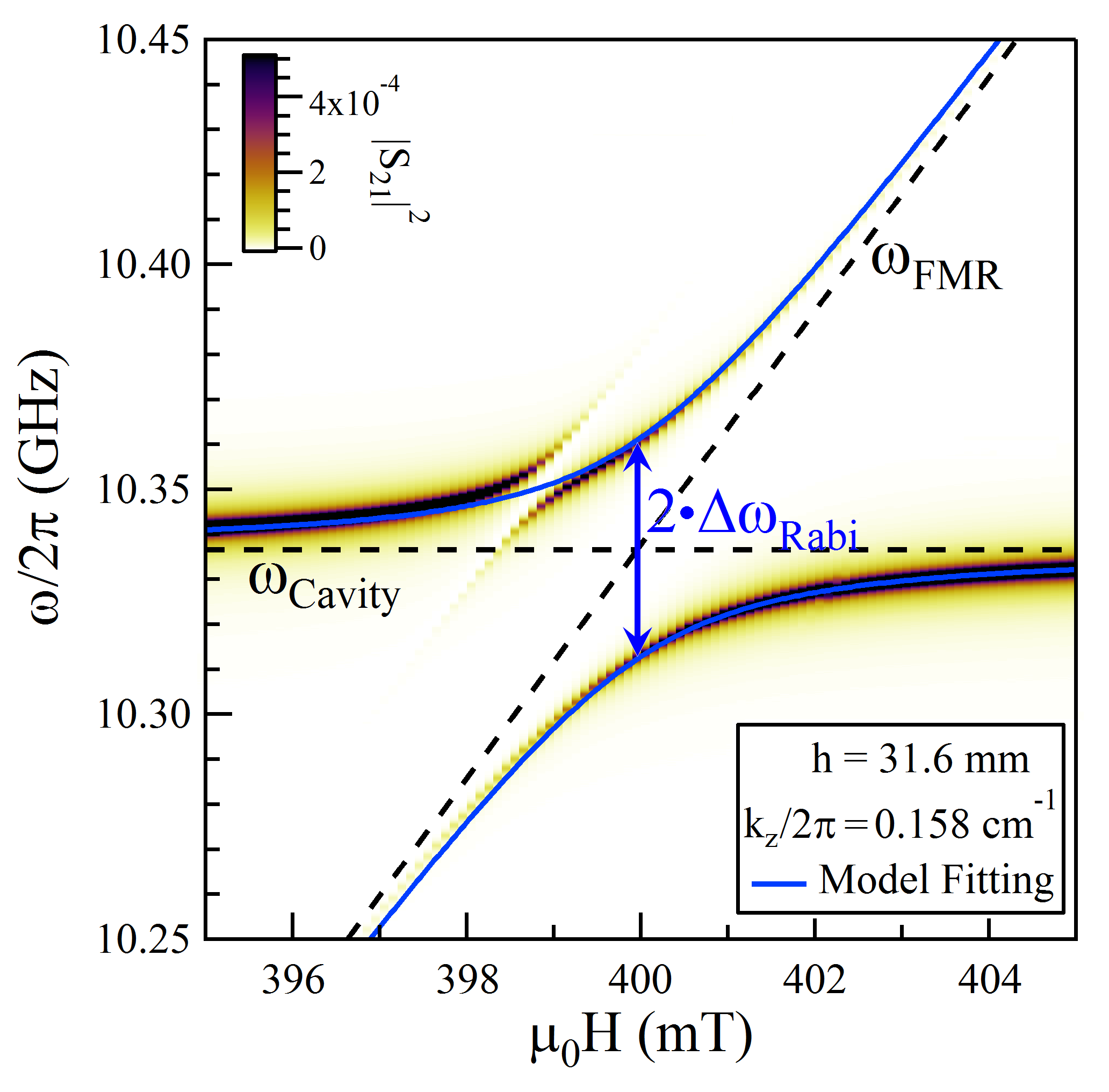}
\caption{(Colour online) An $\omega$-H plot of the microwave transmission, $|S_{21}|^{2}$, through our cavity/YIG system at a cavity height of 31.6 mm (k$_{z}$ = $2\pi\times$ 0.158 cm$^{-1}$). As in Fig. 3(a), the solid blue curve is a fitting of this data to the coupling model in Eq. 1 using $\eta=2.3\times 10^{-5}$, and the black dashed curves indicate the positions of the YIG and cavity resonance modes for the case of no coupling. Here, the Rabi coupling gap, $\Delta\omega_{Rabi}$, can be measured as half the frequency gap between the upper and lower coupled resonance modes at the $\mu_{0}$H value where their uncoupled modes would cross. For equal $\eta$ values, the measured $\Delta\omega_{Rabi}$ will have the same value in $\omega$-H dispersions as in $\omega$-k$_{z}$ dispersions.}
\label{Figure 4}
\end{figure}

Since performing $\omega$-k$_{z}$ dispersion measurements on a magnon-cavity system requires an adjustable cavity design, many studies perform simpler $\omega$-H dispersion measurements to investigate coupling in these systems\cite{A,B,C,F,Q}. Although many of the effects of magnon-cavity coupling will not substantially change between these two dispersion types, the frequency gap, $\Delta\omega_{CMP}$, is not expected to occur in $\omega$-H dispersions. For an $\omega$-H dispersion k$_{z}$ must be kept constant, resulting in a constant cavity resonance frequency, $\omega_{Cavity}$. Inserting this into Eq. 1 gives:

\begin{equation}
(\omega^{2}-\omega_{Cavity}^{2})(\omega-\omega_{FMR})-\eta\omega^{2}\omega_{m}=0
\end{equation}

\begin{figure}[t]
\centering
\includegraphics[width= 7.5 cm]{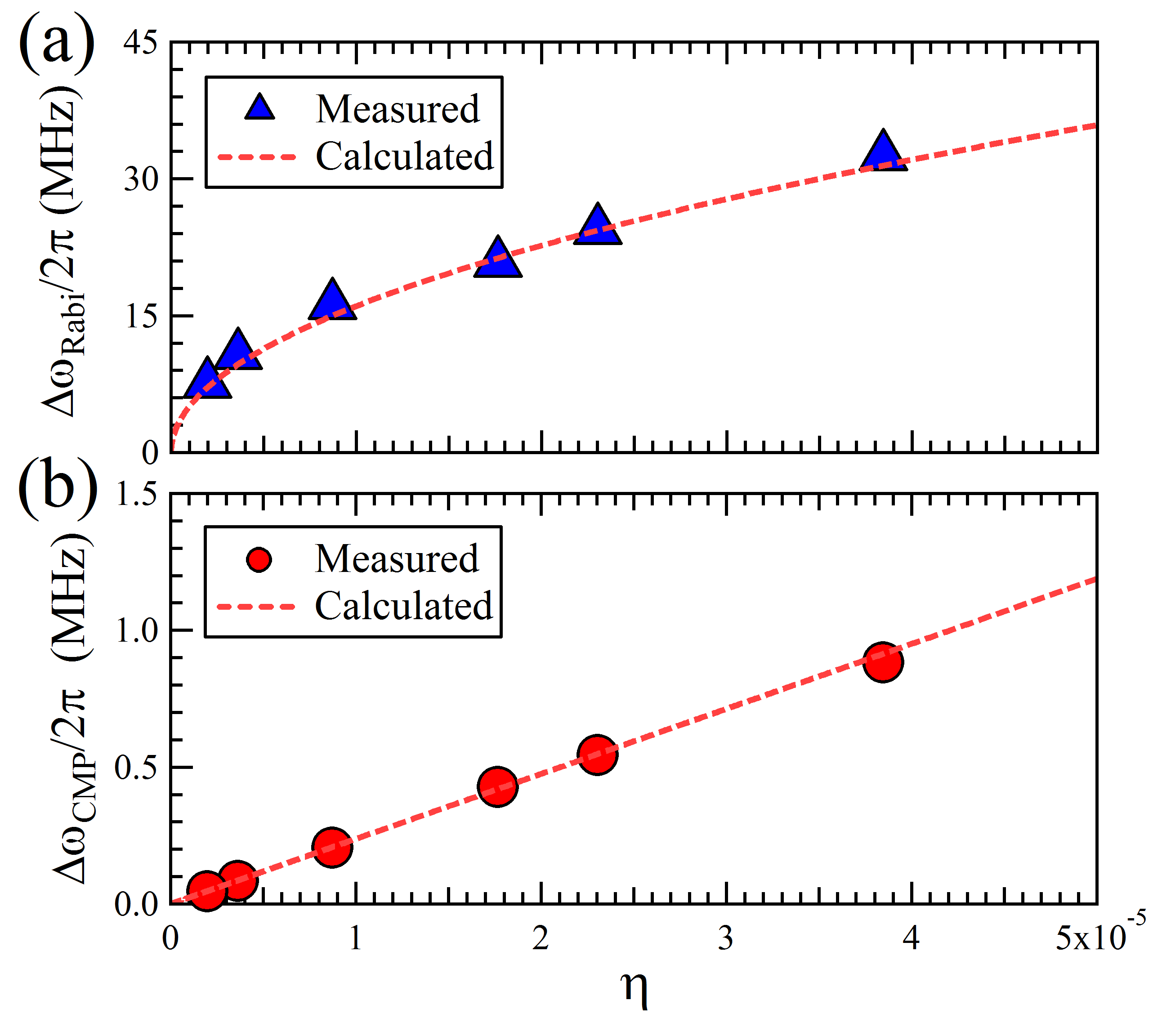}
\caption{(Colour online) \textbf{(a)} A plot of the magnitude of $\Delta\omega_{Rabi}$ (circles), taken from $\omega$-k$_{z}$ measurements at different $\eta$ strengths, compared to the values predicted by Eq. 4 (dashed curve). \textbf{(b)} A plot of the magnitude of $\Delta\omega_{CMP}$ (triangles), taken from $\omega$-k$_{z}$ measurements at different $\eta$ strengths, compared to the values predicted by Eq. 2 (dashed curve).}
\label{Figure 5}
\end{figure}

Looking at this expression, we see that it is equivalent to coupling expressions developed using a quantum model\cite{A,C,F}, equivalent circuit model\cite{F,P,R}, and transfer matrix model\cite{G}. Thus, even with the simplified derivation of $\mu_{eff}$, our model of magnon-cavity coupling is equivalent to these widely accepted models. Fig. 4 shows a measured $\omega$-H transmission dispersion for our coupled cavity/YIG system at k$_{z}$ = $2\pi\times$ 0.158 cm$^{-1}$ (h = 31.6 mm). The resulting plot is similar to the $\omega$-k$_{z}$ dispersion of Fig. 3(a), but now $\omega_{Cavity}$ remains constant while $\omega_{FMR}$ increases linearly with H. Again we see a mode anti-crossing where the two modes approach each other. Here the separation of these two modes at the anti-crossing point will be the same as it was in the $\omega$-k$_{z}$ dispersion of Fig. 3, since at the anti-crossing point the adjustable system parameters (k$_{z}$, H) will be equal. Thus the coupling strength of both dispersion plots can be defined by the Rabi Coupling Gap, $\Delta\omega_{Rabi}$. This coupling gap is produced in multi-mode systems driven by an oscillating field\cite{U}, and can be calculated from Eq. 3, for small $\eta$ values ($\eta$ $<$ 5\%)\cite{N}, as the gap between the two coupled modes when $\omega_{FMR}$ = $\omega_{Cavity}$:

\begin{equation}
\Delta\omega_{Rabi}=\sqrt{\frac{1}{2}\eta\omega_{m}\omega_{Cavity}}
\end{equation}

We can see from Eq. 4 that the magnitude of $\Delta\omega_{Rabi}$ is dependent on the square root of $\eta\omega_{m}$, which represents the total number of net spins averaged over the volume of the cavity, consistent with other model's predictions of the Rabi Gap size. From Eq. 4 we can see that the filling factor, $\eta$, is a key component in determining the magnitude of both $\Delta\omega_{Rabi}$ and $\Delta\omega_{CMP}$. The relations both of these gaps have to $\eta$ is plotted in Fig. 5, where the square root dependence of $\Delta\omega_{Rabi}$ and the linear dependence of $\Delta\omega_{CMP}$ agree well with the measured values from the $\omega$-k$_{z}$ dispersion in Fig 3.

An approximated filling factor allows us to calculate the effective permeability of our coupled magnon-cavity system. Using this effective permeability, magnon-cavity coupling within the system can be studied. The magnitudes of the $\omega$-H Rabi coupling gap, $\Delta\omega_{Rabi}$, and the $\omega$-k$_{z}$ polariton gap, $\Delta\omega_{CMP}$, are both found to be determined by the filling factor used in our model. This indicates that magnon-cavity coupling within the system can be characterized by either of these gap magnitudes, and makes our method of calculating the permeability of the system a useful tool to study coupled magnon-cavity systems. The presence of a mode anti-crossing and the $\Delta\omega_{CMP}$ polariton gap in our $\omega$-k$_{z}$ dispersion measurements show that the features associated with MP coupling remain present in magnon-cavity coupled systems, although with reduced magnitudes reflecting the much smaller material filling factor of our system.

\section{Summary}

In summary, this work clarifies the link between MP and magnon-cavity coupling theories using an MP-based model modified by the effective permeability of a system. Using a height adjustable cavity we produce both $\omega$-k and $\omega$-H dispersion measurements for a magnon-cavity coupled system, and find that the magnitude of the observed mode anti-crossing and polariton gap features of the dispersions are related. The coupling features of both dispersion plots can be reproduced by our model, showing this model to be consistent with, and an extension of, previously used models for MP and magnon-cavity coupling. In both our modelled and measured dispersions, the $\omega$-k and $\omega$-H coupling features are seen to be strongly dependent on the effective permeability of the system. These results show that the effects of magnon-cavity coupling are closely related to those of MP coupling, and that both forms of coupling can be described by a single MP-based model characterized by the effective permeability of a system.

The authors would like to thank B.M. Yao and G. Sawatzky for helpful discussions. P.H. is supported by the UMGF program. M.H. is supported by an NSERC CGSD Scholarship. This work has been funded by NSERC, CFI, and NSFC (No. 11429401) grants (C.-M. Hu).

\end{document}